\documentclass[aps,prl,reprint]{revtex4-1}

\usepackage{blindtext}

\usepackage{graphicx,epsfig,epstopdf}
\usepackage{amsmath}
\usepackage{color}
\usepackage{subfigure}
\usepackage{array}
\usepackage{tabularx}
\usepackage{multirow}
\newcolumntype{L}[1]{>{\raggedright\arraybackslash}p{#1}}
\newcolumntype{C}[1]{>{\centering\arraybackslash}p{#1}}
\newcolumntype{M}[1]{>{\centering\arraybackslash}m{#1}}

\def\_#1{{\bf #1}}
\def\.{\cdot}

\def\Re{{\rm Re\mit}}

\begin{document}

\title{Power-flow Conformal Metamirrors for Engineering Wave Reflections}

\author{
 Ana~D\'{i}az-Rubio$^{1}$, Junfei~Li$^{2}$, Chen~Shen$^{2}$, Steven~A.~Cummer$^{2}$,   and   Sergei~A.~Tretyakov$^{1}$
}
 
\affiliation{$^1$Department of Electronics and Nanoengineering, Aalto University, P.~O.~Box~15500, FI-00076 Aalto, Finland\\
$^2$ Department of Electrical and Computer Engineering, Duke University, Durham, North Carolina 27708, USA}

\begin{abstract} 
Recently, the complexity behind manipulations of reflected fields by metasurfaces has been addressed showing that, even in the  simplest scenarios, non-local response and excitation of auxiliary evanescent fields are required for perfect field control.  Although these solutions theoretically allow to reflect incident plane waves into any desired direction, actual implementations are difficult and, in most cases, require extensive  numerical optimization of the metamirror topology.   In this work we introduce purely local reflective metasurfaces for arbitrary manipulations of the power distribution of reflected waves without excitation of any auxiliary evanescent fields. The reflected fields of such local metamirror contain only the desired propagating waves. 
The method is based on the analysis of the power flow distribution and the adaptation of the reflector shape to the desired distribution of incident and reflected fields. As a result, we find that these power-conformal metamirrors can be easily implemented with conventional passive unit cells.
The results can be used for the design of reflecting surfaces with multiple functionalities and for waves of different physical nature. In this work we present the cases of anomalous reflection and beam splitting, both for acoustic and electromagnetic waves. 
\end{abstract}

\maketitle

\section{Introduction}
Metasurfaces, the two-dimensional versions of metamaterials, have opened new possibilities to control scattering of waves, with many  applications in thin-sheet polarizers, beams splitters, beam steerers, lenses, and more  \cite{yu2014flat,glybovski2016metasurfaces,holloway2012overview}.
The interest in thin structures capable to control and transform impinging waves increased after the formulation of the generalized reflection and refraction law  (GSL) \cite{yu2011light}, which tells that by using small phase-shifting elements  it is possible to control the directions of reflected and transmitted waves.

Among all possible scenarios where metasurfaces can be applied, this work is focused on the analysis of reflective metasurfaces, so-called metamirrors.
In this context, the simplest non-trivial functionality is probably the anomalous reflection, which is the phenomenon of plane-wave reflection in directions different from the specular one.  Anomalous reflection can be obtained by using diffraction gratings (blazed gratings), where the energy scattered into each propagating Floquet harmonic is carefully engineered \cite{bonod2016diffraction,popov1990gratings,kitt2015visible,ra2017meta}. The efficiency of these systems, defined as the percentage of the incident power which is sent into the desired direction,  can be high only if there is not more than one or two unwanted propagating Floquet modes or in the retro-reflection case. Otherwise, there is strong scattering into undesired directions \cite{asadchy2016perfect,estakhri2016wave,diaz2016generalized,epstein2016synthesis,diaz2017perfect,asadchy2017optics}.  

Metasurfaces, which allow subwavelength-scale control of fields, have been proposed as an alternative to conventional gratings, potentially offering full control over the reflection directions. 
Despite the simplicity of this problem, which has been extensively studied for electromagnetic \cite{yu2011light,asadchy2015functional,sun2012high} and acoustic \cite{li2013reflected,zhao2013manipulating,song2016directional,wang2016subwavelength} waves, it was not until recently when the physics of this wave  transformation by metasurfaces was properly understood  \cite{asadchy2016perfect,estakhri2016wave,diaz2016generalized,epstein2016synthesis,diaz2017perfect,asadchy2016perfect,asadchy2017optics}.  
In particular, it was shown that phase-gradient metasurfaces designed based on the generalized reflection law  \cite{yu2011light} can have high efficiencies only if the deflection angle does not exceed 40-45$^\circ$ \cite{diaz2016generalized,asadchy2017optics}. 
Furthermore, notwithstanding much progress in the understanding of anomalous reflective metasurfaces, known methods do not offer means for realizing  more general and complex distributions of fields, where the amplitude, phase and direction of multiple reflected waves could be controlled. The next step towards full engineering of wave reflection is the simultaneous control of two reflected waves. As it was demonstrated in \cite{asadchy2017multi}, flat beam splitting metasurfaces  also require strong non-local response and, consequently, the use of heavy numerical optimizations is inevitable during the design stage.
Finding possibilities for controlling multiple reflected waves without parasitic reflections using local metasurfaces can open new avenues for the design of devices such as holograms or lenses.

To understand the difficulties related to control of reflections from metasurfaces, one can consider power flow in the vicinity of anomalous reflectors. Here, multiple propagating waves with different transverse wavenumbers coexist in one medium, and the interference between them results in inhomogeneous power-flow profiles, where the power flow vector crosses the  metasurface plane. In other words, there will be regions where  power carried by the desired distribution of the incident and reflected waves  ``enters" the metasurface and other regions where it ``emerges" from the surface. It means  that the metasurface   requires periodically distributed  gain/lossy response \cite{estakhri2016wave} or strongly non-local behavior \cite{asadchy2016perfect,diaz2017perfect,diaz2016generalized}. It was shown theoretically that the non-local properties, required for high-efficiency reflections into arbitrary directions, can be in principle realized by excitation of  additional auxiliary evanescent fields  \cite{epstein2016synthesis,ra2017meta} or carefully engineering the surface reactance profile \cite{diaz2017perfect}. The only known experimental realizations of non-local anomalous reflectors are based on extensive numerical optimizations  \cite{diaz2016generalized, asadchy2017optics}, because the intrinsically non-local behavior of any meta-atom combined with the goal to engineer the non-local properties of many interacting meta-atoms complicates the implementation of all non-local solutions. Furthermore, non-local metasurfaces can generate the required set of propagating waves (incident and reflected) only at some distance from the metasurface plane, where the evanescent fields, responsible for the non-local properties, sufficiently decay.

Here we study possibilities to create metamirrors capable of reflecting waves into arbitrary directions without parasitic scattering and without the need to excite any evanescent fields close to the metasurface. In this scenario, the fields in front of the metamirror are perfect combinations of the desired propagating plane waves in the far zone as well as in the near vicinity of the  metasurface. Absence of evanescent fields in front of the metamirror implies that the response is local and that it is possible to design metamirrors using analytical formulas, without any further numerical optimization of complex non-local structures.  
We approach the problem by analysing the distributions of propagating power flow in the desired set of plane waves, not restricting the study to waves of a specific physical nature. 
Previouly, analysis of the power flow distribution has been used for studying surface-relief gratings \cite{popov1990gratings}, where the metallic (or dielectric) shape of the grating can be designed for controlling the energy scattered into a specific diffraction mode. However, these solutions do not ensure exact fulfilment of the boundary conditions on the surface.  The method proposed here allows us to design theoretically perfect anomalous reflectors with rather general functionalities. Illustrations are provided for anomalous reflectors and beam splitters.  The derivations are made for acoustic (the main text) and electromagnetic (Supplementary Materials) scenarios. 

\section{Results}
\subsection{Design methodology}
In this section, we provide a systematic methodology for the design of metamirrors. The approach comprises four steps: (i) Definition of the fields for the desired functionality satisfying the global power balance (all the incident energy is reflected by the metasurface); (ii) Analysis of the power flow distribution and definition of the conformal surface; (iii) Surface impedance calculation; (iv) Implementation with passive elements. 
\subsubsection{Anomalous reflective metamirror}
\begin{figure*}[]
\begin{minipage}{0.5\linewidth}
\subfigure[]{\includegraphics[width = 0.75
\linewidth, keepaspectratio=true]{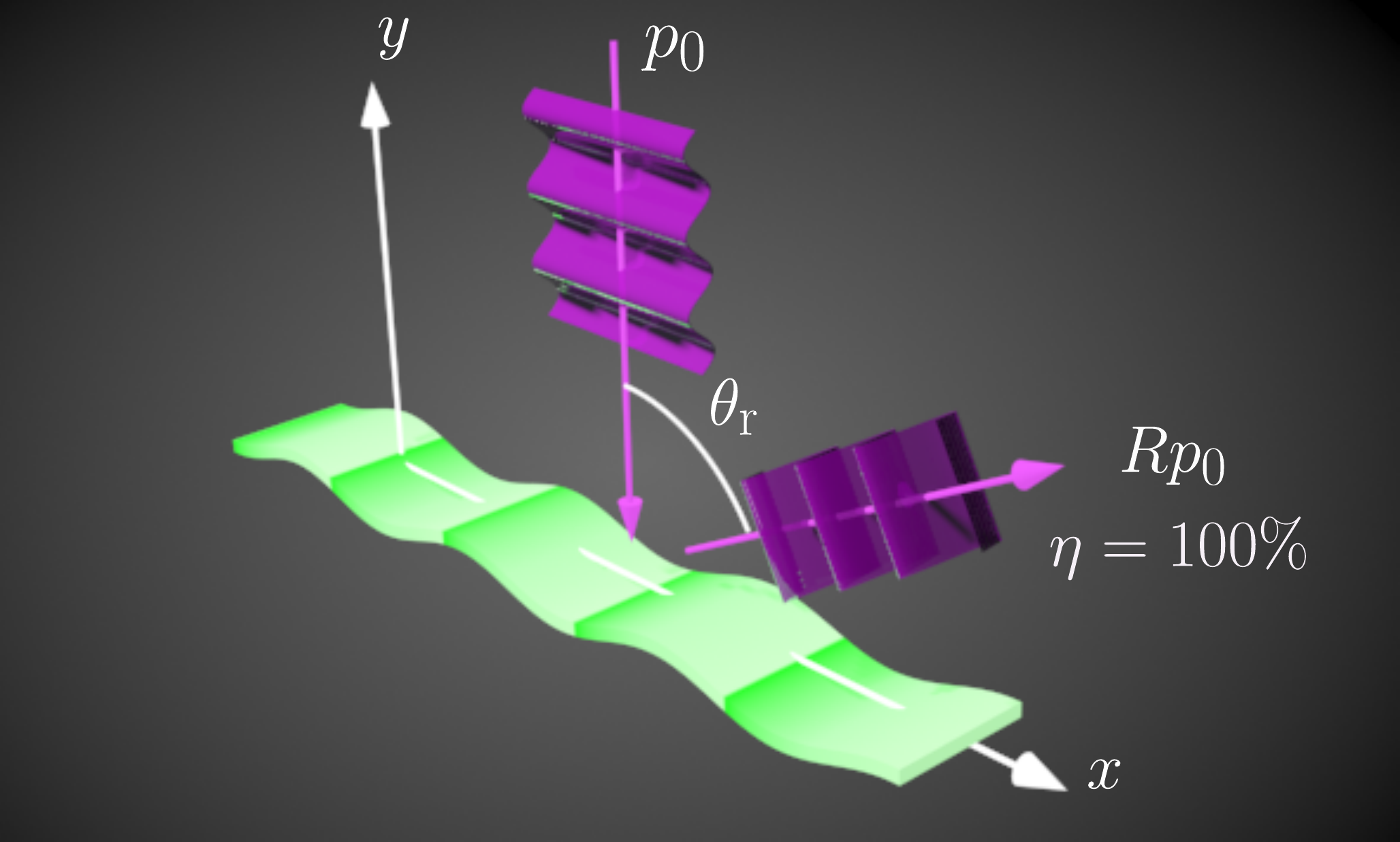}\label{fig:DiazRubioFIG1a}}
\subfigure[]{\includegraphics[width = 0.45\linewidth, keepaspectratio=true]{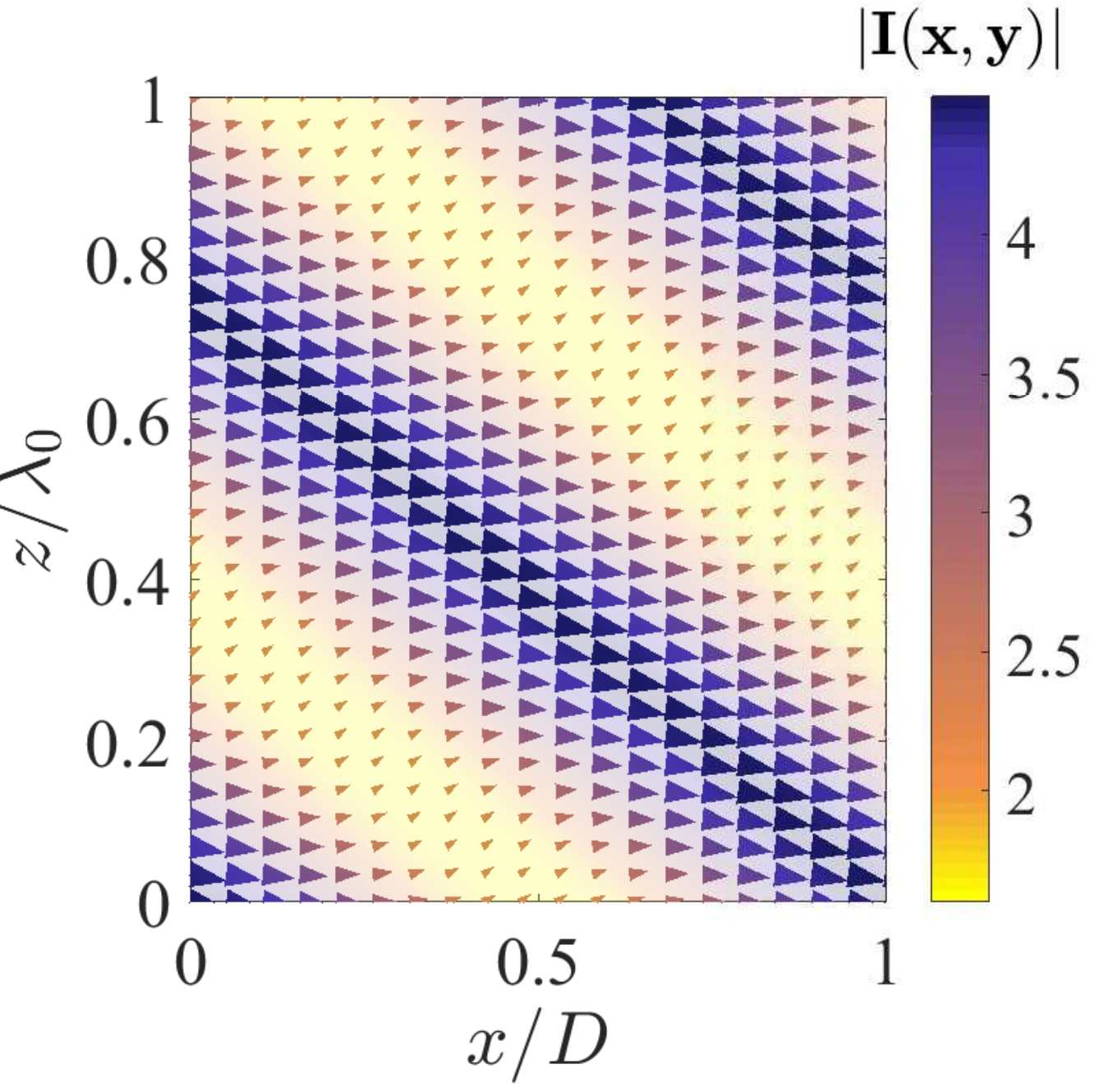}\label{fig:DiazRubioFIG1b}}
	\subfigure[]{\includegraphics[width = 0.45\linewidth, keepaspectratio=true]{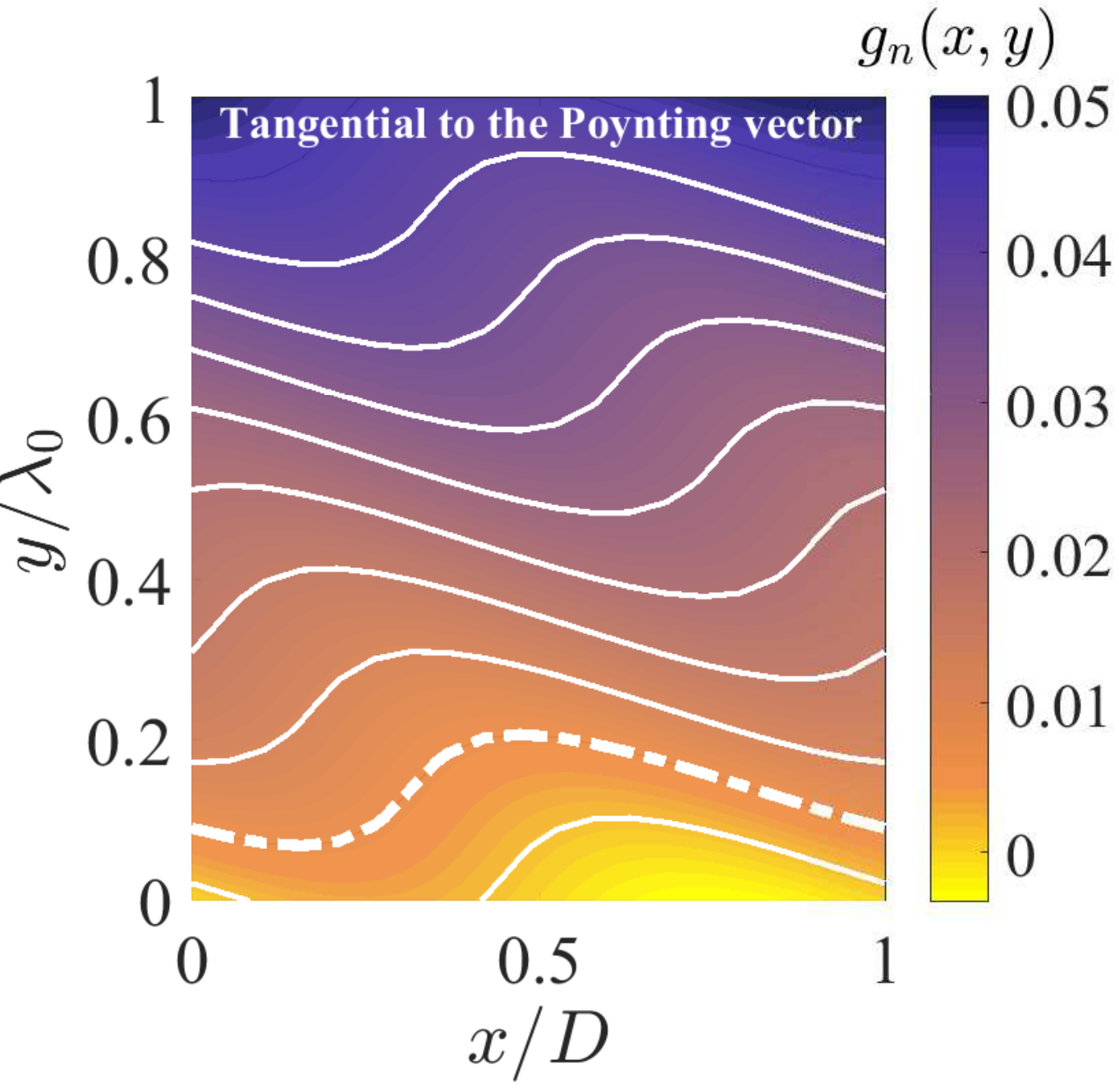}\label{fig:DiazRubioFIG1c}}
\end{minipage}
\begin{minipage}{0.45\linewidth}
	\subfigure[]{\includegraphics[width = 0.8\linewidth, keepaspectratio=true]{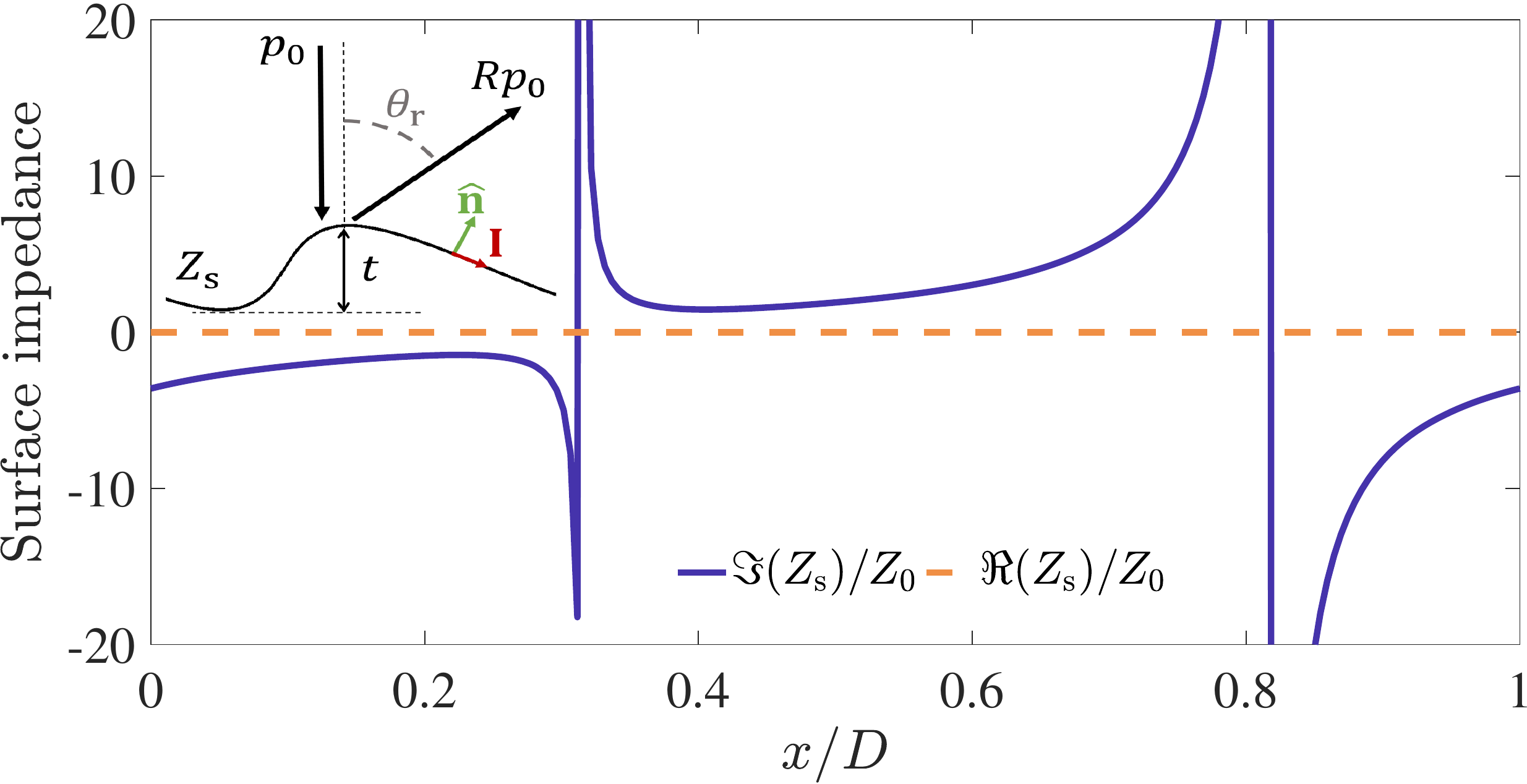}\label{fig:DiazRubioFIG1d}}
	\subfigure[]{\includegraphics[width = .45\linewidth, keepaspectratio=true]{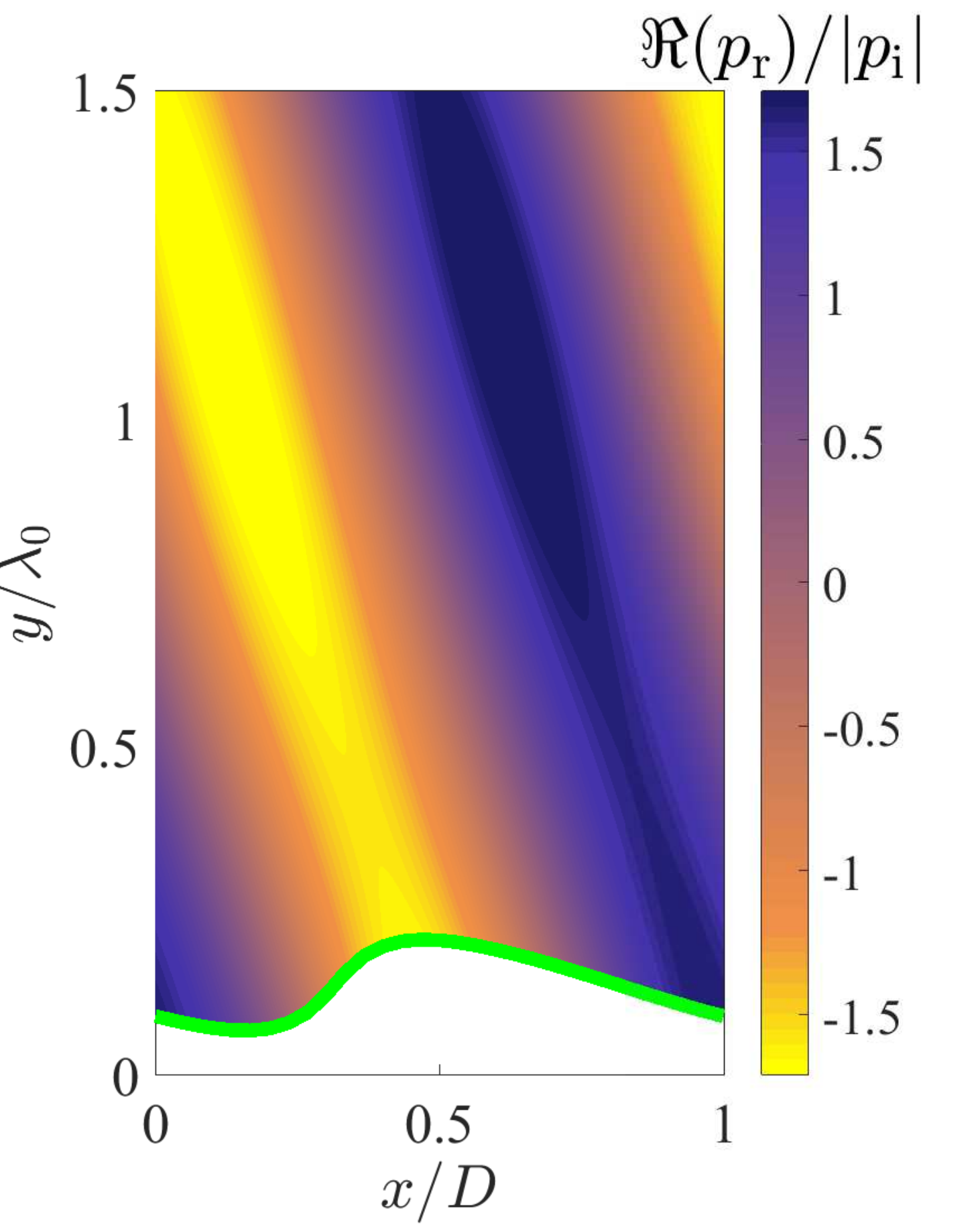}\label{fig:DiazRubioFIG1e}}
	\subfigure[]{\includegraphics[width = .45\linewidth, keepaspectratio=true]{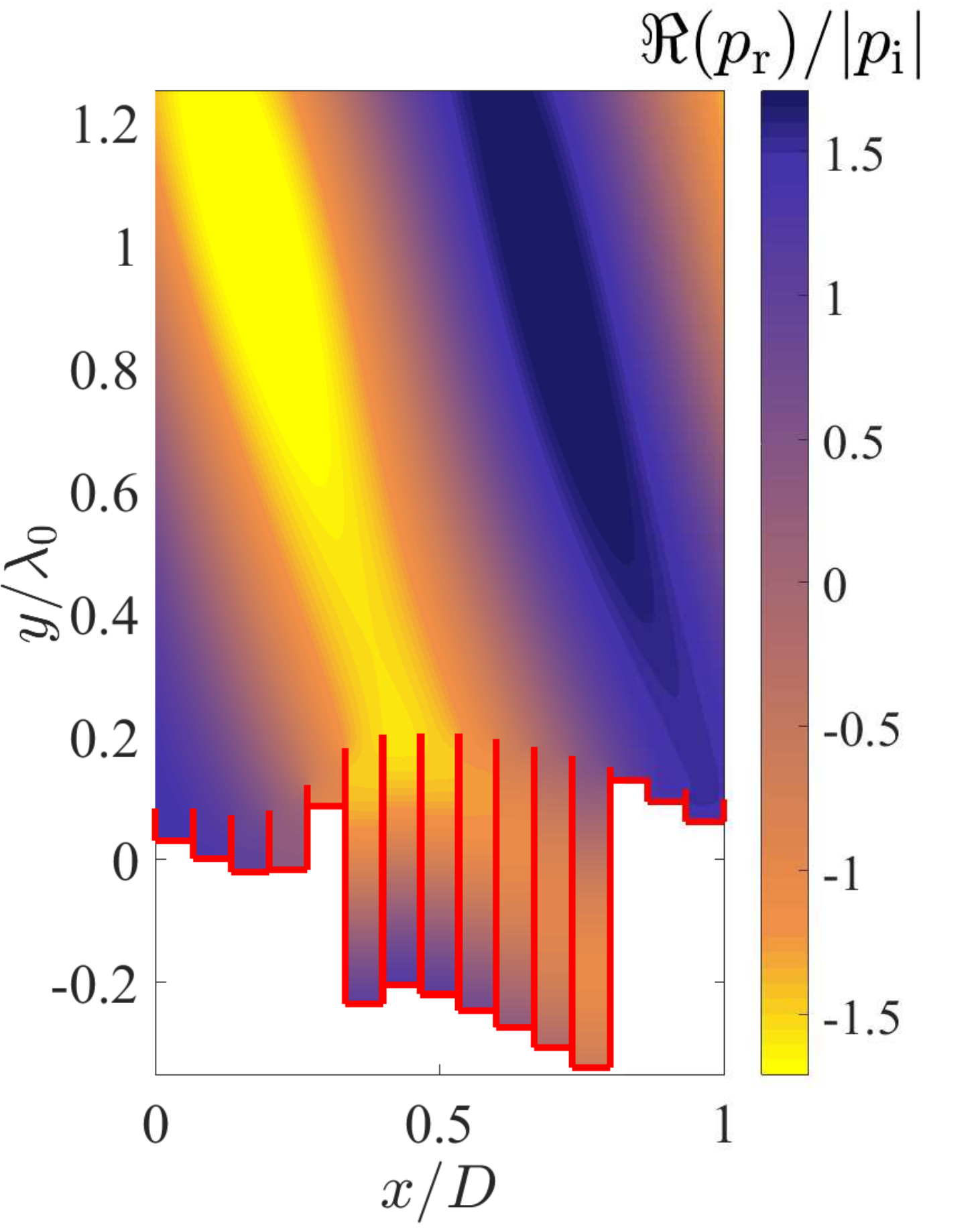}\label{fig:DiazRubioFIG1f}}
\end{minipage}
\caption{\textbf{Anomalous reflective metamirror.}  The study is done for $\phi_{\rm r}=0$, $\theta_{\rm i}=0^\circ$, and $\theta_{\rm r}=70^\circ$. (a) Schematic representation of the problem. (b) Distribution of the intensity vector dictated by  Eqs.~(\ref{eq:Ix}) and (\ref{eq:Iy}). The period of the metasurface can be calculated as $D=\lambda_0/|\sin\theta_{\rm i}-\sin\theta_{\rm r}|$ where   $\lambda_0$ is the wavelength at the operation frequency. (c) The normalized curve level function $g_n(x,y)=g(x,y)/I_0$. White lines represent the level curves, i.e., the curves parallel to the intensity vector at every point. (d) Surface impedance.  The corresponding level curve associated with this impedance is marked with the dashed line in Fig.~\ref{fig:DiazRubioFIG1b}. Numerical simulation of the response of a power-conformal metasurface: (e) Metasurface modeled as an inhomogeneous reactive  boundary. The green line shows the boundary surface. (f) Actual implementation using rigidly ended tubes. The red lines indicate surfaces modeled as hard boundaries. }\label{fig:DiazRubioFIG1}
\end{figure*}

We begin by considering  the anomalous reflection scenario where, requiring absence of any parasitic reflections, we  define an incident sound plane wave and a reflected plane wave with the directions of propagation $\theta_{\rm i}$ and $\theta_{\rm r}$, respectively.  Figure \ref{fig:DiazRubioFIG1a} shows a schematic representation of the problem when $\theta_{\rm i}=0^\circ$.  
Pressure field in this scenario can be expressed as 
\begin{equation}
p(r)=p_0 \left[ e^{-j{\bf k}_{\rm i}\.{\bf r}}+ R  e^{-j{\bf k}_{\rm r}\.{\bf r}}\right],
\end{equation}
where ${\bf k}_{\rm i}=k(\sin \theta_{\rm i}\hat{{\bf x}}-\cos \theta_{\rm i}\hat{{\bf y}})$, ${\bf k}_{\rm r}=k(\sin \theta_{\rm r}\hat{{\bf x}}+\cos \theta_{\rm r}\hat{{\bf y}})$, $k=\omega/c$ is the wave number in the medium, $p_0$ is the amplitude of the incident plane wave, and $R=\vert R \vert e^{j\phi_{\rm r}}$ is the reflection coefficient. The components of the  velocity vector ${\bf v}(r)=v_x(r)\hat{{\bf x}}+v_y(r)\hat{{\bf y}}$ associated with this pressure field read
\begin{eqnarray}
v_x(r)=\frac{p_0}{\eta_0} \left[\sin\theta_{\rm i} e^{-j{\bf k}_{\rm i}\.{\bf r}}+ R \sin\theta_{\rm r} e^{-j{\bf k}_{\rm r}\.{\bf r}}\right],\\
v_y(r)=\frac{p_0}{\eta_0} \left[-\cos\theta_{\rm i} e^{-j{\bf k}_{\rm i}\.{\bf r}}+ R \cos\theta_{\rm r} e^{-j{\bf k}_{\rm r}\.{\bf r}}\right],
\end{eqnarray}
with $\eta_0=c\rho$ being the characteristic impedance of the medium. As it was demonstrated in \cite{asadchy2016perfect,estakhri2016wave,diaz2017perfect}, for ensuring perfect conversion between the incident and the reflected plane waves, avoiding scattering of energy into any other direction, the amplitude of the reflection coefficient has to satisfy  $\vert R \vert=\sqrt{\cos\theta_{\rm i}/\cos\theta_{\rm r}}$.

For proper understanding of the problem we need to examine the intensity vector distribution. The $x$ and $y$ components of the intensity vector can be written as
\begin{equation}
I_x =  I_0 [A + \vert R \vert \left(\sin\theta_{\rm i}+\sin\theta_{\rm r} \right)\cos \left(\Delta{\bf k} \.{\bf r}+\phi_{\rm r}\right)],\label{eq:Ix} 
\end{equation}
\begin{equation}
I_y = I_0\vert R \vert \left(\cos\theta_{\rm r}-\cos\theta_{\rm i} \right)\cos \left(\Delta{\bf k} \.{\bf r} +\phi_{\rm r}\right),
\label{eq:Iy}
\end{equation}
where $A=\sin\theta_{\rm i}+ \vert R \vert^2  \sin\theta_{\rm r}$, $\Delta{\bf k}=k[(\sin\theta_{\rm i}-\sin\theta_{\rm r})\hat{{\bf x}}-(\cos\theta_{\rm i}+\cos\theta_{\rm r})\hat{{\bf y}}]$, and $I_0=\frac{1}{2}\frac{p_0^2}{\eta_0}$.
The first term in Eq.~(\ref{eq:Ix}) can be interpreted as the contributions of the incident and reflected plane waves. The second term describes spatial modulations of the power flow  due to interference of these two waves. Equation~(\ref{eq:Iy}) shows that there is a periodically varying power flow in the normal direction due to interference of the incident and reflected waves. 

Figure~\ref{fig:DiazRubioFIG1a} shows the distribution of the intensity vector when $\phi_{\rm r}=0$,  $\theta_{\rm i}=0^\circ$, and $\theta_{\rm r}=70^\circ$. Detailed inspection reveals that for any horizontal line, for example $y=0$, where one can position a flat metamirror, the intensity vector crosses the surface. This behavior can be described in terms of a complex surface impedance \cite{asadchy2016perfect,diaz2017perfect}, where the real part  takes  positive and negative values, corresponding to ``loss" or ``gain" inside the metamirror. It is worth noticing that the value of the reflection coefficient has been chosen for ensuring the overall power balance between the incident and reflected energies. Thus, loss and  gain compensate each other when averaged over the metasurface period. If the surface is passive and lossless, the periodical modulation of the energy crossing the boundary can be possibly realized arranging some channeling of energy along the metasurface plane, which requires strongly non-local (spatially dispersive) properties. 

Locally responding lossless metasurfaces can be realized only if the real part of the surface impedance is zero, which means that the power is allowed to flow only along the surface without crossing the metasurface boundary.  This condition can be satisfied by defining a specific spatial profile of the metasurface, which would be at all points tangential to the power flow of the desired set of the incident and reflected fields. In this case, energy is not  entering nor emerging from the metasurface.  
To find such spatial profiles, we introduce a vector field which is everywhere  tangential to the power flow. First, we define a vector perpendicular to the intensity vector as ${\bf N}=-I_y\hat{{\bf x}}+I_x\hat{{\bf y}}$. Then we define a scalar function $g(x,y)$ such that $\nabla g(x,y)= {\bf N}$.  In the particular case of anomalous reflection, $g(x,y)$ reads
\begin{equation}
g(x,y)=I_0\left[Ay+B\sin \left(\Delta{\bf k} \.{\bf r}\right)+C\right] , \label{eq:levelcurve}
\end{equation}
where $B=\frac{\vert R \vert }{k}\frac{\cos\theta_{\rm i}-\cos\theta_{\rm r}}{\sin\theta_{\rm i}-\sin\theta_{\rm r}}$ and $C$ is a constant. 
Analysing the spatial distribution of function $g(x,y)$, we identify the  level curves of the function $g(x,y)$, which can be described as $y=f(x)$. Figure~\ref{fig:DiazRubioFIG1b} represents the function $g(x,y)$ and the curves at which it is constant for our example of $\phi_{\rm r}=0$,   $\theta_{\rm i}=0^\circ$, and $\theta_{\rm r}=70^\circ$. At any curve given by Eq.~(\ref{eq:levelcurve}) the power flow is tangential to  this curve. Thus, at these curves we can terminate the field domain by a boundary modeled by a purely imaginary, reactive input impedance. 

To realize a perfectly reflecting metamirror, we select one of such curves and calculate the corresponding impedance. In order to do that, we define the normalized normal vector to such power-conformal metasurface as $\hat{{\bf n}}={\bf N}/|{\bf N}|$ [see inset plot in Fig.~\ref{fig:DiazRubioFIG1d}]. In terms of this vector, the surface impedance is defined as 
\begin{equation}
Z_{\rm s}(x)=\frac{p(x,y_{\rm c})}{-\hat{{\bf n}}\.{\bf v}(x,y_{\rm c})} . \label{eq:impedance_AR}
\end{equation}
This impedance is represented in Fig.~\ref{fig:DiazRubioFIG1d}, where we can see that the real part is indeed identically zero, meaning that a local and lossless design is possible. We have numerically corroborated this finding using a numerical simulation where the metasurface is modeled as a boundary impedance \cite{COMSOL}. The results are reported in Fig.~\ref{fig:DiazRubioFIG1e}, where the scattered pressure is plotted. The green line shows the position of the impedance boundary which models the metasurface. The efficiency of the design is  99\%. It is important to mention that the maximum amplitude (defined as the distance between the maximun and minimun position) of the contour modulation, $t$, is small in terms of the wavelength. 

Thanks to the local, passive and lossless nature of the impedance represented in Fig.~\ref{fig:DiazRubioFIG1d}, we can easily design and realized a curved metamirror which provides the desired response. As a proof of concept we use the simplest phase-shifters, rigidly ended tubes. The input impedance of each tube can be found as $Z_{\rm s, i}=-j\eta_0\cot(k l_i)$ where $l_i$ is the length of each tube. We select the  length of  each  tube according to Eq.~(\ref{eq:impedance_AR}), and this completes the design.  For the particular example of an anomalous reflector for  $\theta_{\rm i}=0^\circ$ and $\theta_{\rm r}=70^\circ$,  Fig.~\ref{fig:DiazRubioFIG1f} shows the scattered pressure of the final design implemented with terminated tubes. Red lines show the tube walls modeled by hard boundary conditions. Specifically, in each period we use 15 tubes with the lengths $0.0524\lambda_0$, $0.0699\lambda_0$, $0.0874\lambda_0$, $0.0961\lambda_0$, $0.0349\lambda_0$, $0.4156\lambda_0$, $0.4068\lambda_0$, $0.4243\lambda_0$, $0.4418\lambda_0$, $0.4563\lambda_0$, $0.4738\lambda_0$, $0.4884\lambda_0$, $0.0029\lambda_0$, $0.0204\lambda_0$, and $0.0349\lambda_0$. The efficiency of the reflector 99\%, without any numerical optimization.

Such simple design based on analytical expressions becomes possible because power-conformal metamirrors do not need excitation and careful engineering of reactive, evanescent fields in the vicinity of the metasurface. Each small portion of the surface responds locally to the fields at its location.  It is important to mention that, in order to reduce the overall thickness of the device, any other phase shifter can potentially be used such as labyrinth-cells \cite{wang2016subwavelength,li2013reflected}, without affecting the performance.  The same approach can be used as a systematic design method for anomalous reflectors for any desired incidence and reflection directions.

\subsubsection{Beam splitting metamirror}

\begin{figure*}[]
\begin{minipage}{0.5\linewidth}
    \raisebox{0mm}{\subfigure[]{\includegraphics[width = 0.8\linewidth, keepaspectratio=true]{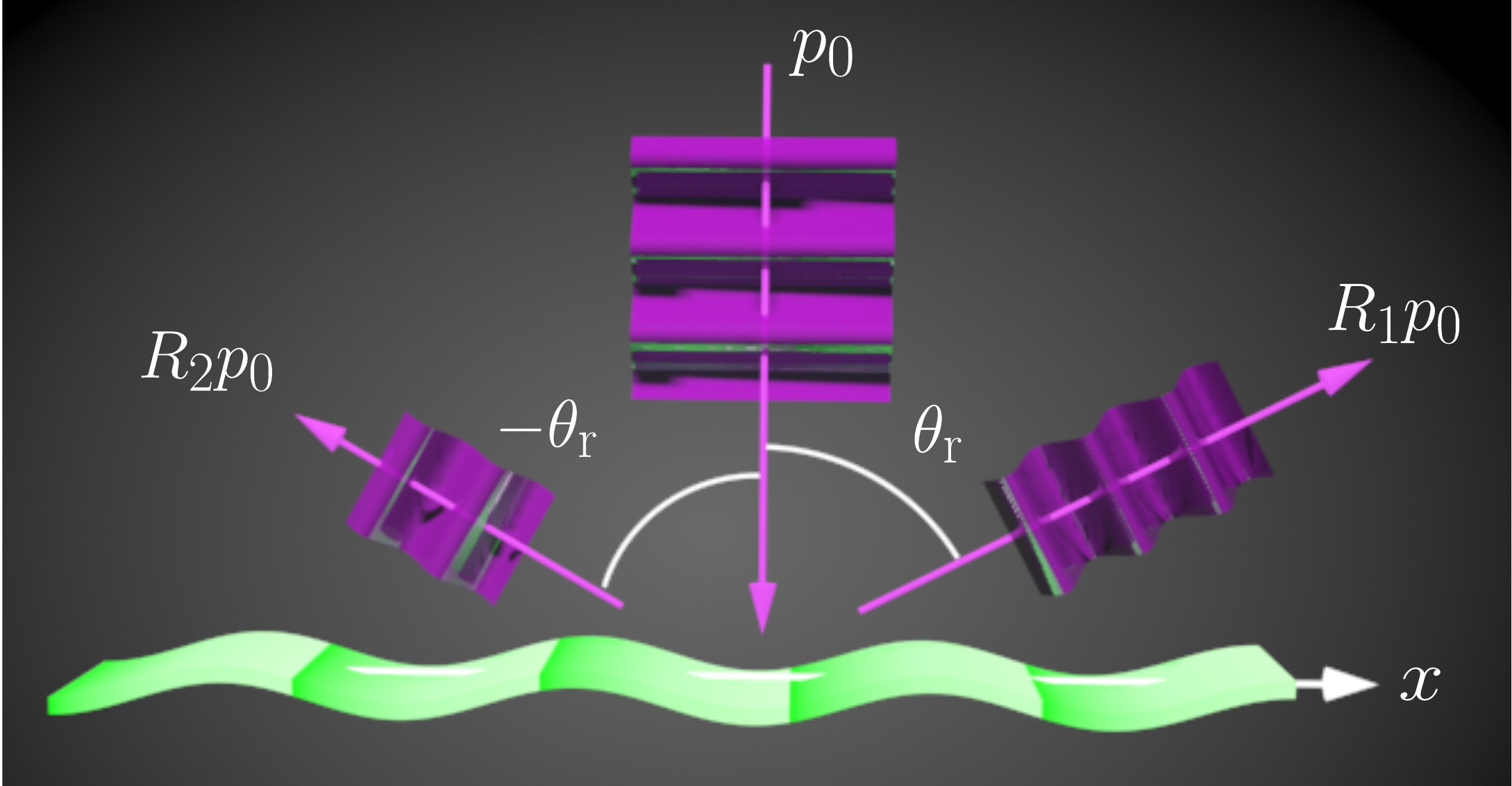}\label{fig:DiazRubioFIG2a}}}
	\subfigure[]{\includegraphics[width = 0.47\linewidth, keepaspectratio=true]{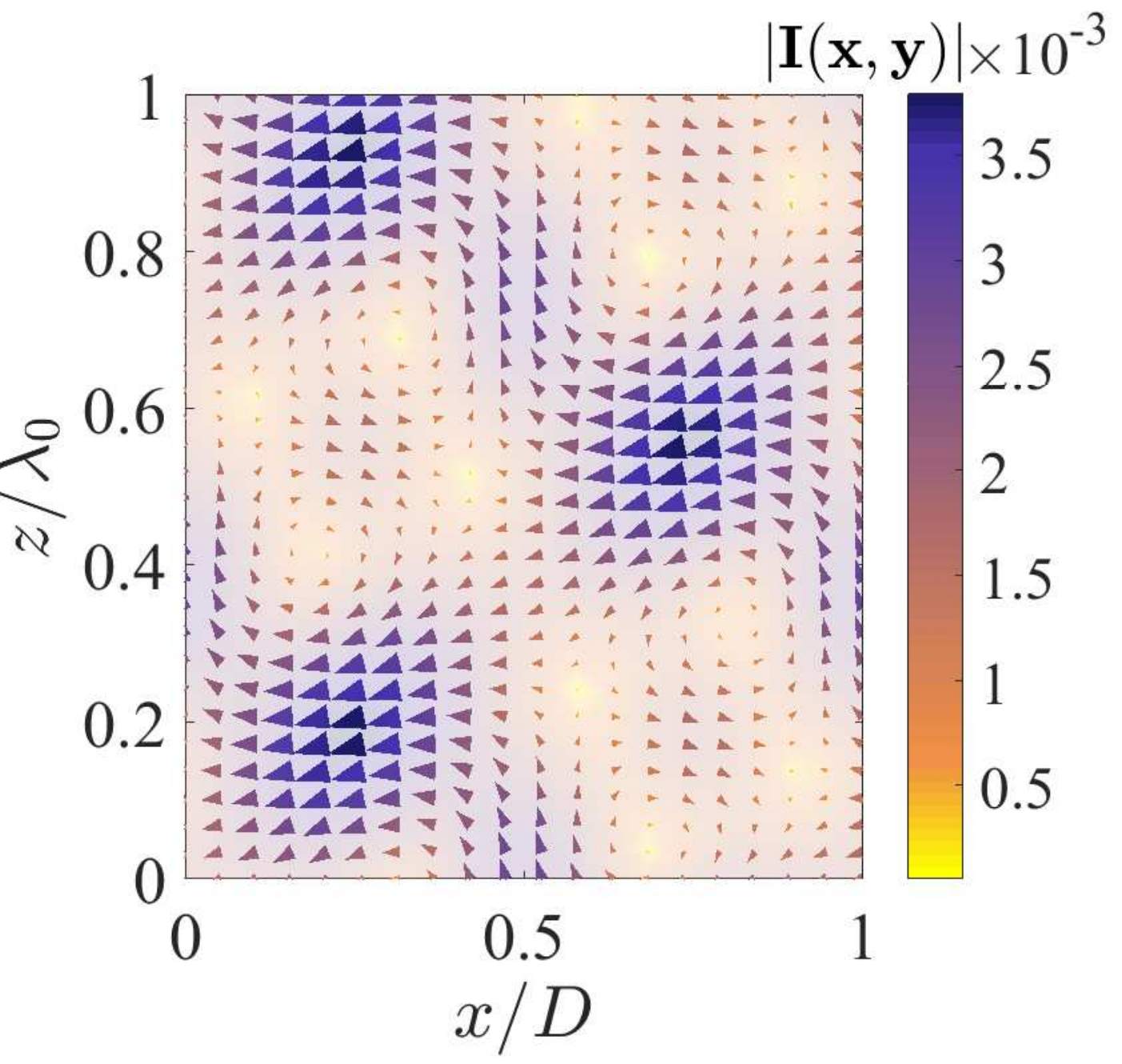}\label{fig:DiazRubioFIG2b}}
	\subfigure[]{\includegraphics[width = 0.49\linewidth, keepaspectratio=true]{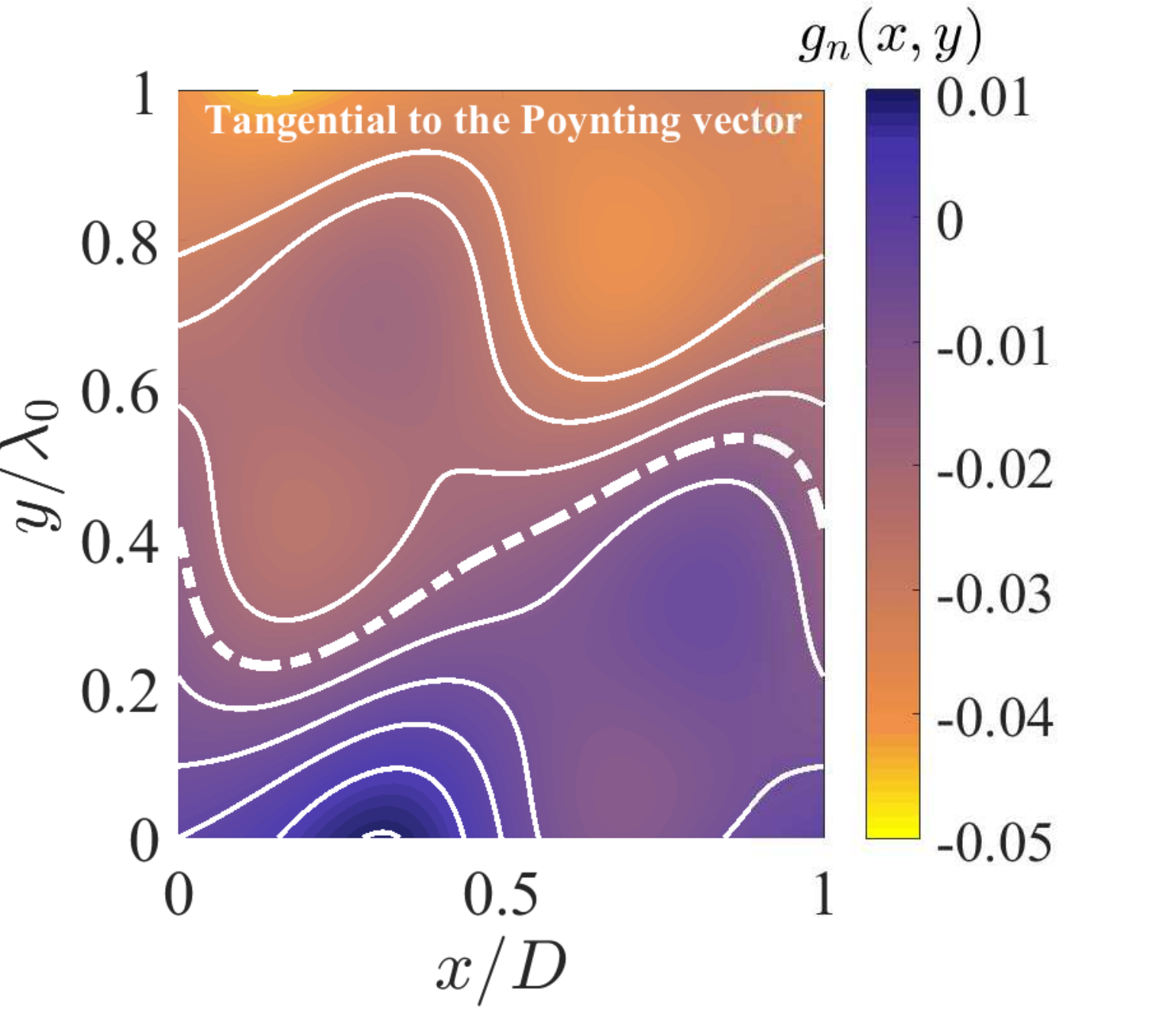}\label{fig:DiazRubioFIG2c}}    
\end{minipage}
\begin{minipage}{0.45\linewidth}
 \subfigure[]{\includegraphics[width = 0.8\linewidth, keepaspectratio=true]{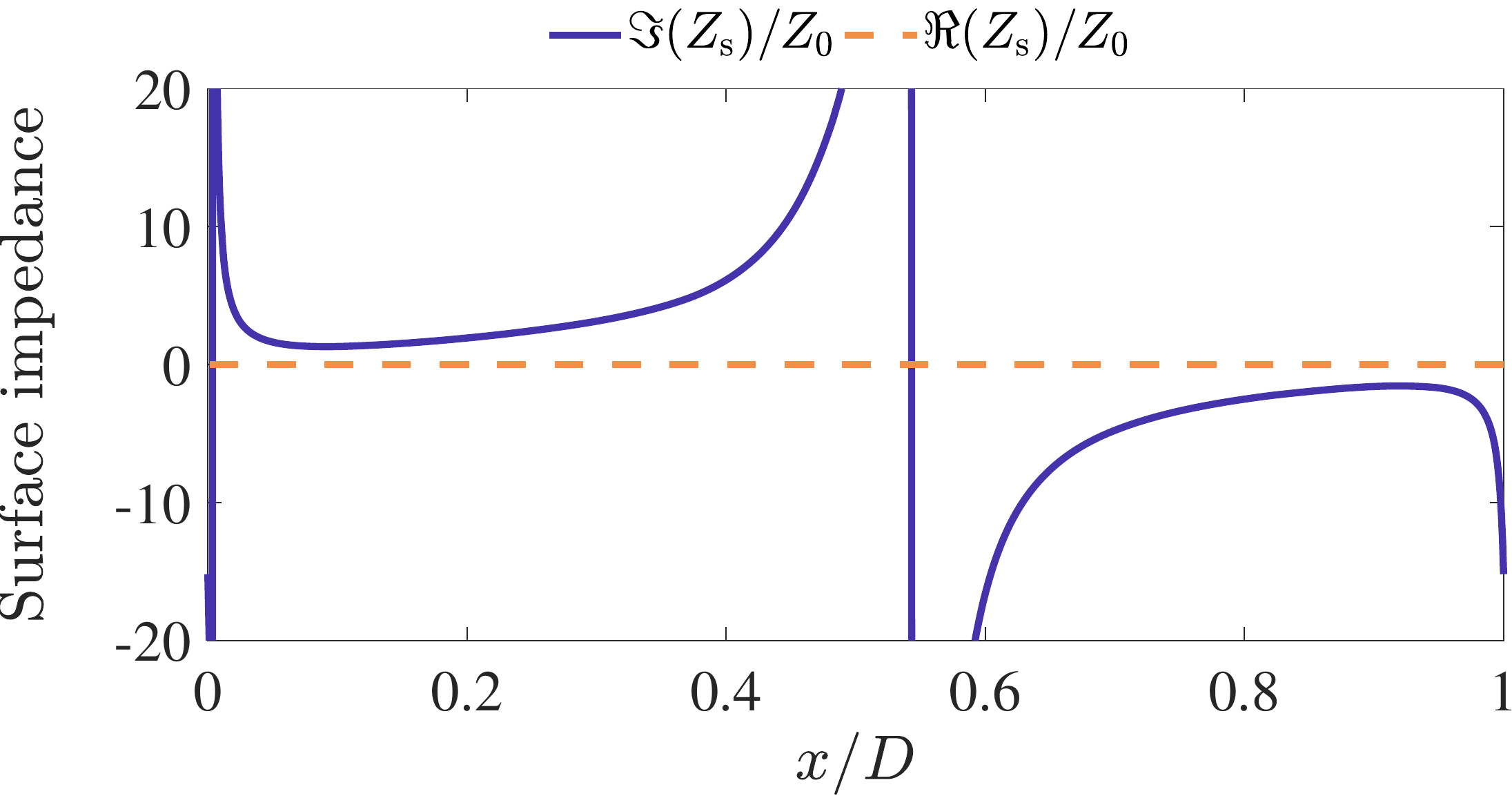}\label{fig:DiazRubioFIG2d}}
\subfigure[]{\includegraphics[width = .47\linewidth, keepaspectratio=true]{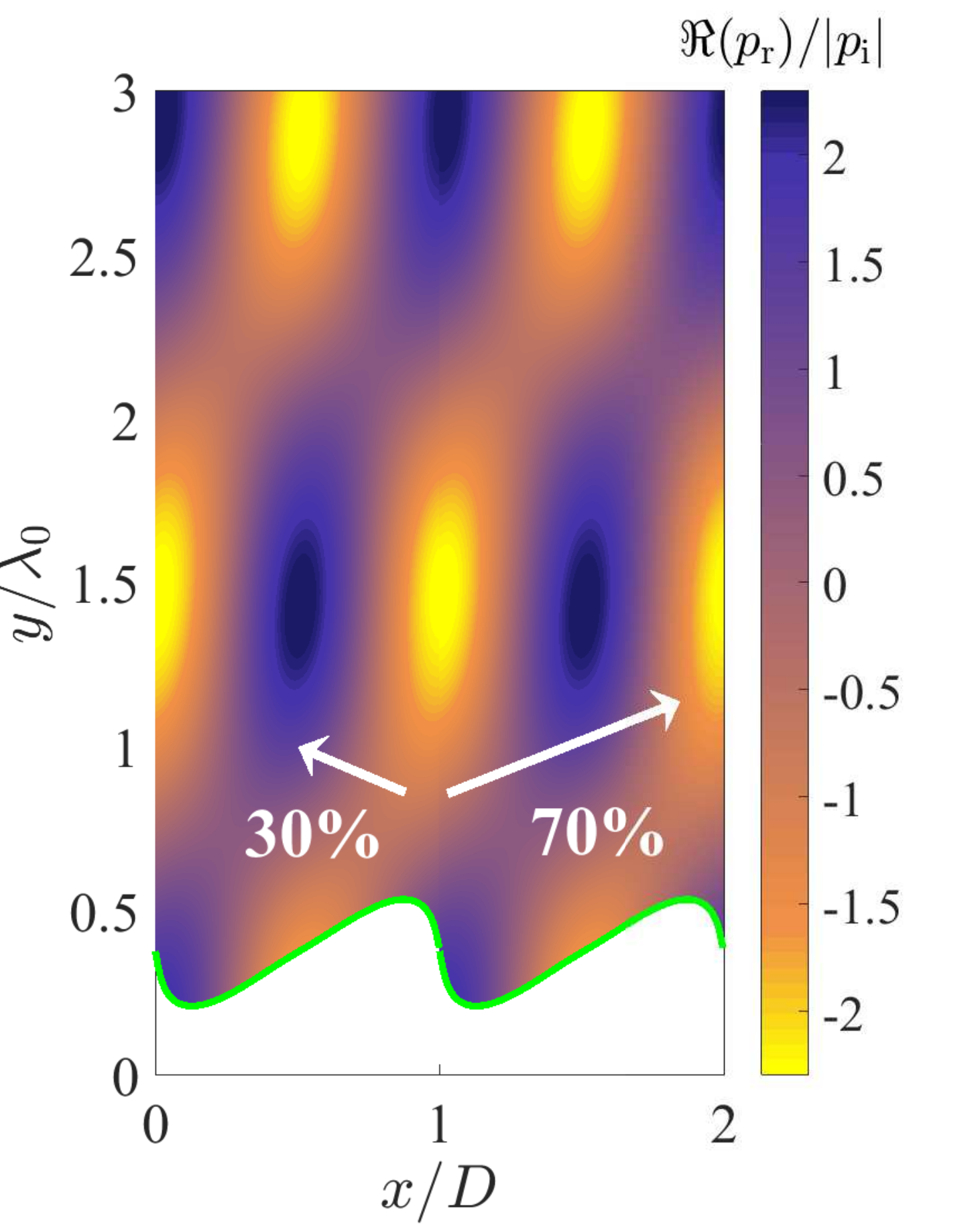}\label{fig:DiazRubioFIG2e}}
 \subfigure[]{\includegraphics[width = .47\linewidth, keepaspectratio=true]{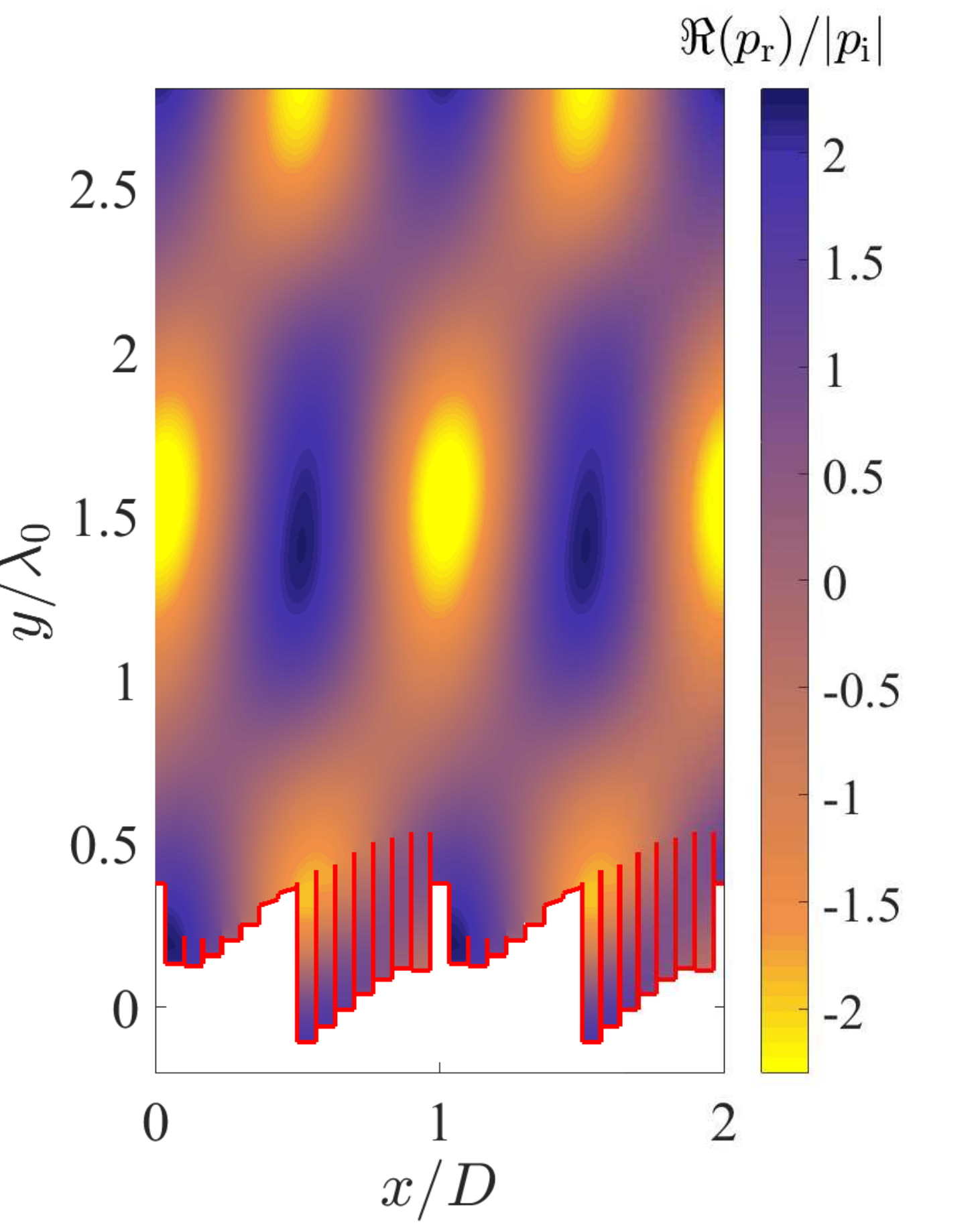}\label{fig:DiazRubioFIG2f}}
\end{minipage}
	\caption{\textbf{Asymmetric beam splitter (70\%  and 30\%)}. The analysis is done for $\phi_1=\phi_2=0$, $\theta_{\rm i}=0^\circ$, and $\theta_{\rm r}=\pm 70^\circ$. (a) Schematic representation of the problem.  (b) Distribution of the intensity. The period of the metasurface equals $D=\lambda_0/|\sin\theta_{\rm i}-\sin\theta_{\rm r}|$, where   $\lambda_0$ is the wavelength at the operation frequency. (c) The normalized curve level function $g_n(x,y)=g(x,y)/I_0$. White lines represent the level curves, i.e the curves parallel to the intensity vector. (d) Surface impedance. The corresponding level curve associated with this impedance is marked with the dashed  line in Fig.~\ref{fig:DiazRubioFIG2b}. Numerical results for the power-conformal metasurface: (e) Metasurface modeled as an impedance  boundary. The green line shows the position of the boundary. (f) Actual implementation using rigidly ended tubes. Red lines show tube walls modeled as hard boundaries. }\label{fig:DiazRubioFIG2}
\end{figure*}

The introduced method can be used for creation of more complex field distributions and for other functionalities. Here we provide an example of a metasurface capable of splitting waves coming from a certain direction into two reflected waves, propagating along two different desired directions. As it was shown in \cite{asadchy2017multi}, this functionality also requires non-local response or additional evanescent fields. In this case, the pressure field can be expressed as 
\begin{equation}
p(r)=p_0 \left[ e^{-j{\bf k}_{\rm i}\.{\bf r}}+ R_1  e^{-j{\bf k}_{\rm r1}\.{\bf r}}+ R_2  e^{-j{\bf k}_{\rm r2}\.{\bf r}}\right],
\end{equation}
where $R_1=\lvert R_1\rvert e^{j\phi_1}$ and $R_2=\lvert R_2\rvert e^{j\phi_2}$  represent the relative complex amplitudes of the reflected waves. As an example, we assume that the metasurface is illuminated normally, $\theta_{\rm i}=0^\circ$, and the reflected beams are sent into $\pm\theta_{\rm r}$ [see Fig.~\ref{fig:DiazRubioFIG2a}]. In this case, the corresponding wavenumbers read  ${\bf k}_{\rm i}=k\hat{{\bf y}}$, ${\bf k}_{\rm r1}=k(\sin \theta_{\rm r}\hat{{\bf x}}+\cos \theta_{\rm r}\hat{{\bf y}})$, and ${\bf k}_{\rm r2}=k(-\sin \theta_{\rm r}\hat{{\bf x}}+\cos \theta_{\rm r}\hat{{\bf y}})$.  This notation allows us to model and design not only symmetric splitters where the incident power  is equally divided  between the two reflected waves, but realize any other distribution of power  between the two waves which fulfills the power conservation condition  $(\lvert R_1\rvert^2+\lvert R_2\rvert^2)\cos\theta_{\rm r}=1$. As it has been shown in \cite{asadchy2016perfect}, flat metasurfaces for implementing this functionality also require  strong non-local response. Our aim here is to find a local, passive, and lossless realization by using a power-flow conformal metamirror. Following the same approach as above, we need to find  a surface profile $y=f(x)$ where the  corresponding surface impedance $Z_{\rm s}$ is purely imaginary. 

First we find a suitable surface which is tangential to the power flow in the desired set of three plane waves. In this case, the intensity distribution ${\bf I}(x,y)=\frac{1}{2}[\Re(p v_x^*)\hat{{\bf x}}+\Re(p v_y^*)\hat{{\bf y}}]$ depends on the reflection angle $\theta_{\rm r}$ and on the amplitudes of the reflected waves $R_1$ and $R_2$. As an example, we design a metamirror which sends  70\%  and 30\% of the incident power into $\pm 70^\circ$ and $\phi_1=\phi_2=0$. The corresponding amplitudes of the reflection coefficients are $R_1=1.43$ and $R_2=0.94$.  The power flow distribution for this case is represented in Fig.~\ref{fig:DiazRubioFIG2b}, where we clearly see the intensity modulations produced by interfering incident and reflected  waves. The function whose level curves will define the tangential contours to the intensity vector  can be expressed  as 
\begin{equation}
\begin{split}
&g(x,y) =I_0G(x)+I_0F(y)+\\
&I_0\frac{\cos\theta_{\rm r}-1}{k\sin\theta_{\rm r}}\left[R_1\sin \left(\Delta{\bf k}^- \.{\bf r}\right)-
R_2\sin \left(\Delta{\bf k}^+ \.{\bf r}\right)\right],
\end{split}
\label{eq:g2}
\end{equation}
where $\Delta{\bf k}^\pm=k[\pm\sin\theta_{\rm r}\hat{{\bf x}}-(1+\cos\theta_{\rm r})\hat{{\bf y}}]$ measures the intensity modulation strength.  
The expressions for functions  $G(x)$ and $F(y)$ can be written as
\begin{equation}
 F(y)=(R_1^2-R_2^2)\sin\theta_{\rm r}y
\end{equation}
\begin{equation}
\begin{split}
G(x) = [1-(R_1^2+R_2^2)\cos\theta_{\rm r}]x-\frac{R_1R_2}{k}\frac{\cos\theta_{\rm r}}{\sin\theta_{\rm r}}\sin(2k_xx),\end{split}
\end{equation}
where $k_x=k\sin\theta_{\rm r}$. Function $g(x,y)$ is plotted in  Fig.~\ref{fig:DiazRubioFIG2c}. Now we can define possible profiles of local metamirrors, which are shown by white lines. Among all the possible surfaces we chose the one marked with the dashed line.  We can see that the amplitude of the surface modulation is larger than in the anomalous reflective metamirror: $t=0.3\lambda$. The impedance associated with this curve is presented in Fig.~\ref{fig:DiazRubioFIG2d}. 

Figure~\ref{fig:DiazRubioFIG2e} shows the real part of the scattered field obtained with numerical simulations where the metasurface is modeled as a reactive impedance boundary. The field map  shows the interface pattern of plane waves.  The amplitude of the reflection coefficients in this numerical study are  $R_1=1.43$ and $R_2=0.92$. This result is in agreement with the design criteria. For the actual implementation we can use the same configuration where the desired impedance is fulfilled by rigidly ended tubes of different lengths. Figure~\ref{fig:DiazRubioFIG2f} shows the results of a numerical simulation of an actual structure which produces the desired response. The two reflected waves carry $70\%$ and $29\%$ of the incident power. As in any other metasurface  design, discretization of the ideally continuous surface is an important issue. We need to ensure that the impedance profile is smoothly implemented by an array of discrete phase-shifters. Small discrepancy is caused by discretization of the ideally continuous surface. 

\subsection{Experimental verification}

\begin{figure*}[]
\begin{minipage}{0.4\linewidth}
\subfigure[]{\includegraphics[width = 1
\linewidth, keepaspectratio=true]{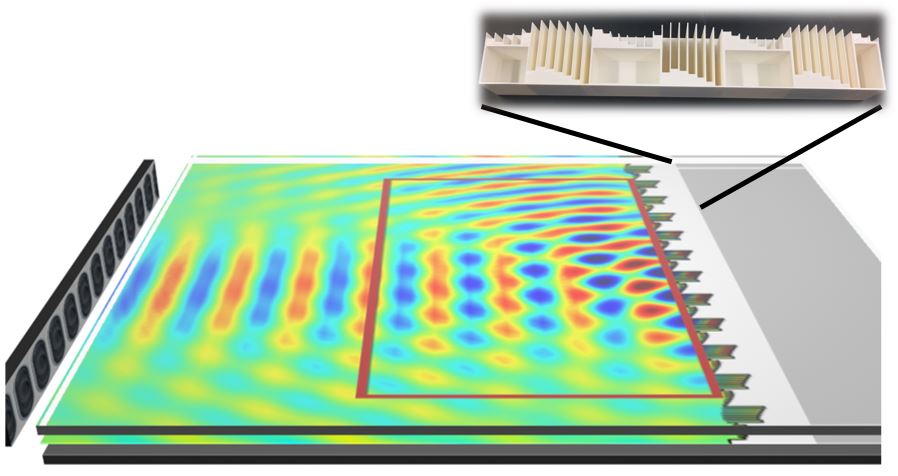}\label{fig:DiazRubioFIG3a}}
\subfigure[]{\includegraphics[width = 0.95
\linewidth, keepaspectratio=true]{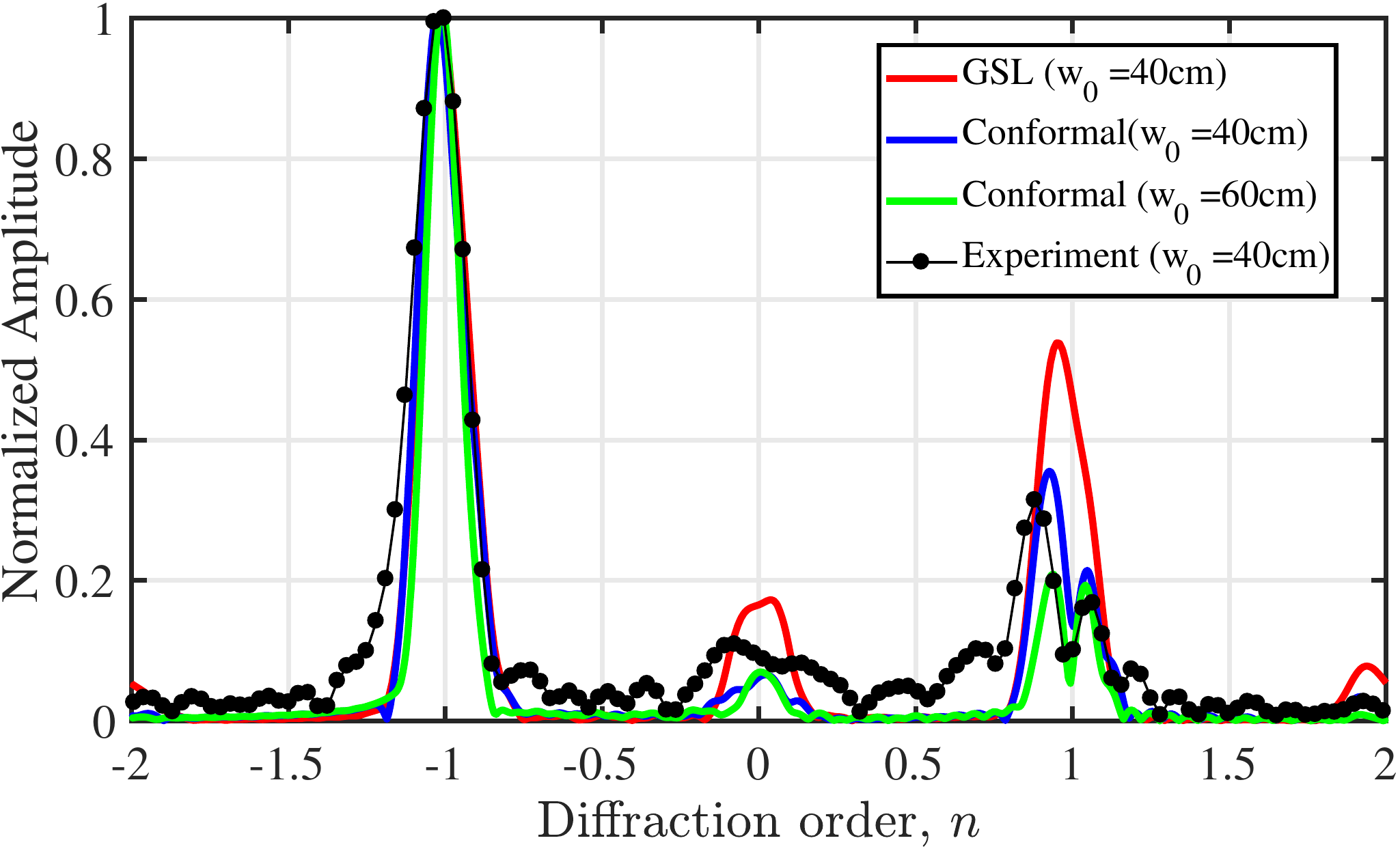}\label{fig:DiazRubioFIG3b}}
\end{minipage}
\begin{minipage}{0.55\linewidth}
\subfigure[]{\includegraphics[width = 1
\linewidth, keepaspectratio=true]{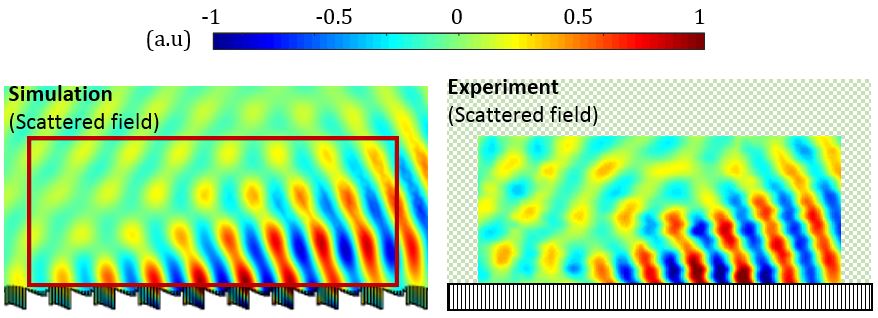}\label{fig:DiazRubioFIG3c}}
\subfigure[]{\includegraphics[width = 1
\linewidth, keepaspectratio=true]{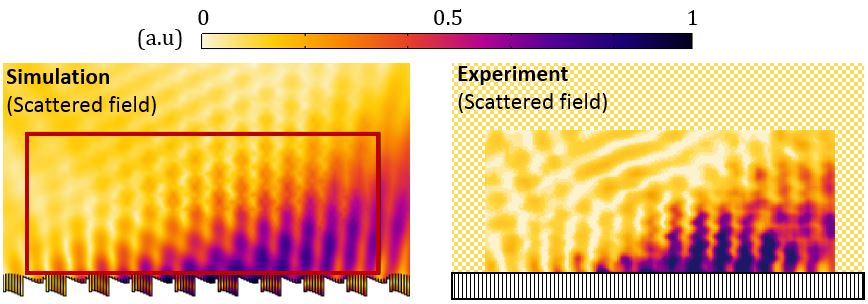}\label{fig:DiazRubioFIG3d}}
\end{minipage}
\caption{\textbf{Experimental verification.} (a) Schematic representation of the experimental setup and a photograph of the fabricated sample. (b) Comparison between the normalized scattering of the anomalous reflective metamirror (experimental and numerical) and a GSL design. (c-d) Analysis of the real part (c) and magnitude square (d) of the experimental pressure field and the comparison with the numerical simulations.}\label{fig:DiazRubioFIG3}
\end{figure*}
The theory is then verified with experiments. As a proof-of-concept demonstration, we choose an acoustic metamirror capable of reflecting normally incident acoustic waves into the $70^\circ$ direction. The metamirror is composed of 3D printed closed-end tubes, where the surface geometry follows the conformal contour describing the power flow direction, as illustrated in Fig.~\ref{fig:DiazRubioFIG1f}. The operational frequency is chosen to be 3000~Hz, and the width of each tube is 8 mm, smaller than 0.1~$\lambda$.The length of a period of the fabricated sample is 12 cm, with thickness being 6.3 cm, around half of the operational wavelength. A photo of one segment of the fabricated sambe with 3 period is shown Fig.~\ref{fig:DiazRubioFIG3a}. The final sample consists of 12 periods.  

 In the experimental verification, a spatially modulated Gaussian beam is used for illuminating the sample (see Methods for more details about the simulation and the experimental beam generation). For obtaining the scattered fields, two measurements are performed. First, the sample is placed in the setup and the total field is acquired, i.e. the sum of incident and scattered fields. The incident field and parasitic scattering from the setup are removed by subtracting the fields measured in the absence of the sample. Left panels of Figs.~\ref{fig:DiazRubioFIG3c} and ~\ref{fig:DiazRubioFIG3d} show the real part and magnitude of the scattered fields by the metamirror when the width of the beam is $w_0=40$~cm. In this results we can clearly see that more energy goes into the desired direction, however, there is a residual amount of energy scattered into other directions. This imperfection is a consequence of the finite width of the beam.
It is important to notice that the sample has been designed for a plane wave to plane wave transformation and it is not optimized the transformation of beams, so for wider beams the metamirror efficiency is higher. For a deeper analysis of this feature, one can compare the performance of the metamirror when it is illuminated with different beams. Specifically, we compare the response when the beam width is 40~and~60cm. The efficiency is further analyzed by performing the Fourier transform on the fields along the line exiting the metamirror, and the results are shown Fig.~\ref{fig:DiazRubioFIG3b}. From this analysis, we can see that the energy scatterted into undesired direction is dramatically reduced when the width of the beam increases.  For comparison purposes, the analysis of a GSL-based metamirror implemented with the same number of elements is also included. We can conclude that the efficiency of the conformal metamirror is higher than the corresponding conventional design.

The sample is secured in a 2D waveguide for field mapping, the detailed experimental setup is described in Methods. Figs.~\ref{fig:DiazRubioFIG3c} and ~\ref{fig:DiazRubioFIG3d} show the simulated and measured acoustic fields at 3000 Hz. Excellent agreement can be observed and it can be seen clearly that the reflected field contains mainly the $70^\circ$ wave component. The small discrepancies may be attributed to non-perfect Gaussian beam generation, fabrication errors, and inevitable dissipation loss. The Fourier analysis result is shown in Fig. ~\ref{fig:DiazRubioFIG3b} where we can confirm the agreement with the simulations. In both simulations and experiments, almost all of the energy is localized at $k_x=\sqrt[]{3}/2 k_0$, which is the desired direction of the outgoing wave. The measured efficiency of the metamirror is $96.9\%$, which validates our approach (see Methods for more details about this calculation).       

\section{Discussion}
In this paper, we have introduced a multi-physics design method for creation of acoustic or electromagnetic metamirrors for general shaping of reflected waves. Examples of anomalous reflectors and beam splitters have been provided. In contrast to known anomalous reflectors, the proposed local, passive, and lossless structures ensure theoretically perfect performance for arbitrary deflection angles, extending the range of accessible functionalities of both diffraction gratings and phase-gradient reflective metasurfaces. It is important to stress that the introduced design approach does not need  any numerical optimizations, offering full physical insight into complex reflection and diffraction phenomena and giving a clear advantage in device design.  
Conformal metasurfaces have been used to create cloaking devices, optical or acoustic illusions, and lenses. In all these examples, conformal metasurfaces are thought to adapt to the \emph{shape} of   scattering or reflecting bodies. Here we have proposed a concept of conformal metasurfaces which adapt to the desired power distribution of the \emph{fields}. Since this concept is applicable in all  scenarios where the gradient of the desired field structure is continuous, it can be used to realize various complex field transformations, such as focusing or beam shaping. The experimental validation reported in this work is the first implementation of an anomalous reflective acoustic metamirror which overcomes the efficiency limitations of GSL-based designs.

\section{Methods}
\subsection{Numerical simulations}
The simulations were performed with the commercial finite
element analysis solver COMSOL Multiphysics. 
The infinite systems are modeled by one period using Floquet periodic conditions. The simulation shown in Fig.~\ref{fig:DiazRubioFIG1e} and Fig.~\ref{fig:DiazRubioFIG2e} are calculated with \textit{Impedance Boundaries} and defining the values according to impedances represented in Fig.~\ref{fig:DiazRubioFIG1d} and Fig.~\ref{fig:DiazRubioFIG2d}. The simulation of the proposed designs [see Fig.~\ref{fig:DiazRubioFIG1f} and Fig.~\ref{fig:DiazRubioFIG2f} ] are calculated with \textit{Sound Hard Boundary} conditions. In these simulations, the illumination is a perfect plane wave implemented with  \textit{Background Pressure Field} domain condition.

For the simulations of the experiment, we use a finite number of periods and Gaussian beam illumination. The Gaussian beam propagating in y-direction is expressed as
\begin{equation}
p_{\rm i}=p_0\frac{w_0}{w(y)}e^{\frac{-x^2}{w(y)^2}}e^{-jk\frac{x^2}{2R(y)}}e^{jk(y-y_{0})}e^{-j\eta(y)},
\end{equation}
where $p_0$ is the beam amplitude, $w_0$ is the spot radius, $w(y)=w_0 \ \sqrt[]{1+(\frac{y-y_0}{y_{\rm R}})^2}$ defines the spot size variation as a function of the distance from the beam waist, $y_{\rm R}=\pi w_0^2/\lambda$ is the Rayleigh range, $R(y)=(y-y_0)[1+(\frac{y_{\rm R}}{y-y_0})^2])$ is the curvature radius, and $\eta(y)=\arctan(\frac{y-y_{0}}{y_{\rm R}})$ is the phase change close to the beam waist. The boundaries of the metasurface are set as hard walls. The background media is modelled as a semi-circle with radius 1.2~m and \textit{Plane Wave Radiation } conditions. The excitation is implemented with  \textit{Background Pressure Field} domain condition. The wall of the metasurface are modelled as \textit{Sound Hard Boundary} conditions.

\subsection{Field mapping measurements}
The samples under test were fabricated with fused deposition modeling (FDM) 3D~printing where the printed material is acrylonitrile butadiene styrene (ABS) plastic with density of 1180~$\rm{kg/m^3}$ and speed of sound 2700~m/s. The walls are considered to be acoustically rigid since the characteristic impedance of the material is much larger than that of air. A loudspeaker array with 28 speakers sends a Gaussian modulated beam normally to the metasurface and the field is scanned using a moving microphone at a step of 2~cm. The acoustic field at each spot is then calculated using Fourier Transform. The reflected field is calculated by filtering out the incident using 2D Fourier transform. The overall scanned area is 100 by 40~cm and the signal at each position is averaged out of four measurements to reduce noise.

\subsection{Measurement of the efficiency}

The efficiency of the metasurface when the metal surface is illuminated by a Gaussian beam cannot be extracted directly from the amplitude of the reflected beam. Due to the multiple wavenumbers associated with the finite size beam this amplitude can be distorted. For an accurate calculation of the efficiency, we use the Fourier transform of the pressure fields along a line over the metasurface [see Figure \ref{fig:DiazRubioFIG3b}].

This analysis gives the amplitude of all the Fourier components. However, for calculating the efficiency we only use the amplitudes of the $n=-1,0,1$ harmonics which correspond to the propagating waves at $70^\circ$, $0^\circ$, and $-70^\circ$. The power carried by each component is calculated as $P_n=A_n^2\cos\theta_n$ where $A_n$ is the amplitude of the $n$-harmonic and $\theta_n$ defines the direction of propagation. Finally, the efficiency of the metasurface can be calculated as 
\begin{equation}
\eta=\frac{P_{-1}}{\sum_{n=-1,0,1}P_n}.
\end{equation}
It is important to notice that in this definition of the efficiency the dissipation losses are not included.
\section*{Acknowledgements}

This work was supported by the Academy of Finland (projects 287894  and 309421) and by the Multidisciplinary University Research Initiative grant from the Office of Naval Research (N00014-13-1-0631).

\bibliographystyle{apsrev4-1}
\bibliography{references}



\end{document}